\documentclass[ aip,
 amsmath,amssymb,
 reprint,%
]{revtex4-1}
\usepackage[utf8]{inputenc}
\usepackage{amsmath}
\usepackage{amssymb}
\usepackage{graphicx}
\usepackage{breqn}
\usepackage{mathtools}
\usepackage{amsthm}
\usepackage{color}
\usepackage{breqn}
\usepackage[hidelinks]{hyperref}
\usepackage{etoolbox}
\usepackage{svg}
\usepackage[nodisplayskipstretch]{setspace}
\usepackage{float}
\usepackage{epstopdf}
\usepackage[justification=centering]{caption}
\usepackage{cancel}

\makeatletter
\preto\maketitle{%
  \begingroup\lccode`~=`,
  \lowercase{\endgroup
  \let\saved@breqn@active@comma~
  \let~}\active@comma 
}
\appto\maketitle{%
  \begingroup\lccode`~=`,
  \lowercase{\endgroup
  \let~}\saved@breqn@active@comma 
}

 \@ifundefined{textcolor}{}
 {%
   \definecolor{BLACK}{gray}{0}
   \definecolor{WHITE}{gray}{1}
   \definecolor{RED}{rgb}{1,0,0}
   \definecolor{GREEN}{rgb}{0,1,0}
   \definecolor{BLUE}{rgb}{0,0,1}
   \definecolor{CYAN}{cmyk}{1,0,0,0}
   \definecolor{MAGENTA}{cmyk}{0,1,0,0}
   \definecolor{YELLOW}{cmyk}{0,0,1,0}
 }

\usepackage{bm}
\makeatother

\def\be{\begin{eqnarray}}
\def\ee{\end{eqnarray}}

\definecolor{JOT-color}{named}{blue}
\definecolor{CSF-color}{named}{orange}

\begin{document}
\title{Role of the absorption on the spin-orbit interactions of light with Si nano-particles}

\author{Jorge Olmos-Trigo}
\email{jolmostrigo@gmail.com}
\affiliation{Donostia International Physics Center (DIPC),  20018 Donostia-San Sebasti\'{a}n, Basque Country, Spain}

\author{Cristina Sanz-Fern\'andez}
\affiliation{Centro de F\'{i}sica de Materiales (CFM-MPC), Centro Mixto CSIC-UPV/EHU,  20018 Donostia-San Sebasti\'{a}n, Basque Country, Spain} 

\author{Diego R. Abujetas}
\affiliation{Donostia International Physics Center (DIPC),  20018 Donostia-San Sebasti\'{a}n, Basque Country, Spain}
\affiliation{Instituto de Estructura de la Materia (IEM-CSIC), Consejo Superior de Investigaciones Cient\'{\i}ficas, Serrano 121, 28006 Madrid, Spain}

\author{Aitzol Garc\'{i}a-Etxarri}
\affiliation{Donostia International Physics Center (DIPC),  20018 Donostia-San Sebasti\'{a}n, Basque Country, Spain}
\affiliation{Centro de F\'{i}sica de Materiales (CFM-MPC), Centro Mixto CSIC-UPV/EHU,  20018 Donostia-San Sebasti\'{a}n, Basque Country, Spain}

\author{Gabriel Molina-Terriza}
\affiliation{Donostia International Physics Center (DIPC),  20018 Donostia-San Sebasti\'{a}n, Basque Country, Spain}
\affiliation{Centro de F\'{i}sica de Materiales (CFM-MPC), Centro Mixto CSIC-UPV/EHU,  20018 Donostia-San Sebasti\'{a}n, Basque Country, Spain}
\affiliation{IKERBASQUE, Basque Foundation for Science, 48013 Bilbao, Basque Country, Spain}

\author{Jos\'e A. S\'anchez-Gil}
\affiliation{Instituto de Estructura de la Materia (IEM-CSIC), Consejo Superior de Investigaciones Cient\'{\i}ficas, Serrano 121, 28006 Madrid, Spain}

\author{F. Sebastián Bergeret}
\affiliation{Donostia International Physics Center (DIPC),  20018 Donostia-San Sebasti\'{a}n, Basque Country, Spain}
\affiliation{Centro de F\'{i}sica de Materiales (CFM-MPC), Centro Mixto CSIC-UPV/EHU,  20018 Donostia-San Sebasti\'{a}n, Basque Country, Spain}

\author{Juan José S\'aenz}
\affiliation{Donostia International Physics Center (DIPC),  20018 Donostia-San Sebasti\'{a}n, Basque Country, Spain}
\affiliation{IKERBASQUE, Basque Foundation for Science, 48013 Bilbao, Basque Country, Spain}

\begin{abstract}
The conservation of the photon total angular momentum  in the incident direction in an axially symmetric scattering process  is a very well known fact. Nonetheless, the re-distribution of this conserved magnitude into its spin and orbital components, an effect known as the spin-orbit interaction (SOI) of light, is still a matter of active research. Here, we discuss the effect of the absorption on the SOI in the scattering of a sub-wavelength Silicon particle. Describing the scattering process of a electric and magnetic dipole, we show via the asymmetry parameter that the
SOI of light in the scattering of high refractive index nanoparticles endures in the presence of optical losses. This effect results in optical mirages whose maximum values surpass those of an electric dipolar scatterer. 
\end{abstract}


\maketitle    
\section{Introduction}

Spin-orbit interactions (SOI) of light  have attracted an increasing interest in the recent years\citep{bliokh2015spin}.
In particular, this re-distribution of the angular momentum \cite{allen2003optical} (AM) into its spin (SAM) and orbital (OAM) contributions \cite{he1995direct,moe1977conservation,crichton2000measurable,berry2005orbital, marrucci2006optical,bliokh2011spin,mansuripur2011spin,marston2018humblet}, 
has been 
thoroughly studied in absorptionless targets\citep{schwartz2006backscattered,araneda2018wavelength}.
Moreover, it has been shown that this re-distribution per photon induces after scattering a shift on the apparent location of the targets 
\cite{onoda2004hall,bliokh2006conservation,hosten2008observation,berry2011lateral}
and, specifically, on spherical scatterers. In the latter case, the scattering of plane waves with well-defined helicity, $\sigma = \pm 1$, on scatterers described by a single polarizability \citep{arnoldus2004dipole,arnoldus2008subwavelength} as well as on larger multipolar spherical scatterers \citep{haefner2009spin}, has been studied.
These apparent displacements are referred to as SOI optical mirages \citep{Olmos2018SOI,olmos2019enhanced}. 
In particular, for high refractive index nano-particles with electric and magnetic dipolar responses \cite{evlyukhin2010optical,Garcia-Etxarri2011,kuznetsov2016optically},
the SOI can lead to enhanced optical mirages 
\cite{Olmos2018SOI,olmos2019enhanced,gao2018enhanced,shi2019enhanced}
at the so-called first Kerker condition \cite{kerker1983electromagnetic,nieto2011angle,gomez2011electric,geffrin2012magnetic,Person2013,fu2013directional},
where  scattering preserves duality symmetry \cite{zambrana2013duality,zambrana2013dual} and  the maximum allowed exchange of AM per photon from spin to orbital contributions ($s_z = -\sigma$ and $\ell_z = 2\sigma$, respectively)  leads to a diverging optical mirage in the backscattering direction\citep{olmos2018asymmetry}.

In all previous works, the absence of optical losses guaranteed the conservation of the total AM, i.e., the total number of photons.
However, in absorbing systems, the incoming radiation induces a net optical torque on the target \citep{marston1984radiation,nieto2015optical} and the SOI of light remain unexplored.

In this work, we discuss the effects of absorption on the SOI of light in the scattering of an electric and magnetic dipolar Silicon (Si) nano-particle whose optical properties have attracted a large interest  \cite{evlyukhin2010optical,Garcia-Etxarri2011,kuznetsov2016optically,kuznetsov2012magnetic,luk2015optimum,bag2018transverse}
As a simple consequence of the scatterer axial symmetry, based on the conservation of the AM per photon in the incident direction, $j_z = \ell_z + s_z = \sigma$, we demonstrate that an exotic ``super-momentum'' regime can be achieved, where $|\ell_z| > |j_z|$ \citep{bliokh2017optical}. We explicitly show that in this regime, even with absorption, an enhanced optical mirage can be generated with respect to the pure dipolar case. 
By analyzing the changes in the far-field polarization, it is possible to retrieve fine information about the scattering particle \cite{rodriguez2010optical,nechayev2018weak,neugebauer2018weak,olmos2018asymmetry} and, as we will show, the SOI optical mirage can
be described in terms of the degree of circular polarization (DoCP) in the far field limit. 
 

\section{Theory}

The scattering from an arbitrary multipolar sphere can be conveniently expressed in the helicity basis \citep{fernandez2012helicity,fernandez2013electromagnetic} as
\begin{align}\label{SO:1}
{\bf{E}}_\sigma^{\rm{scat}} = {\bf{E}}_{\sigma +} + {\bf{E}}_{\sigma -}, && \text{with} && \hat{\bm{\Lambda}}{\bf{E}}_{\sigma \pm} = \pm  {\bf{E}}_{\sigma \pm},
\end{align}
where the sub-indexes $+$ and $-$ refer to  positive and negative  eigenvalues of the helicity operator for monochromatic waves, $\hat{\bm{\Lambda}} = (1/k) \bm{\nabla} \times$.

By applying the helicity operator to the scattered fields ({Eq. \ref{SO:1}}), the helicity density after scattering, or in other words, the DoCP, can easily be obtained,
\begin{equation}\label{SO:2}
\Lambda = \frac{{{\bf{E}}_\sigma^{\rm{scat*}}} \cdot \left(  \hat{\bm{\Lambda}} {\bf{E}}^{\rm{scat}}_\sigma \right)}{{{\bf{E}}_\sigma^{\rm{scat}}}^* \cdot {\bf{E}}^{\rm{scat}}_\sigma} = \frac{|{\bf{E}}_{\sigma +}|^2 - |{\bf{E}}_{\sigma -}|^2}{|{\bf{E}}_{\sigma +}|^2 + |{\bf{E}}_{\sigma -}|^2} = \frac{V}{I},
\end{equation}
where $V$ and $I$ { are} two of the so-called 
Stokes parameters. 

\begin{figure}
\includegraphics[width=1\columnwidth]{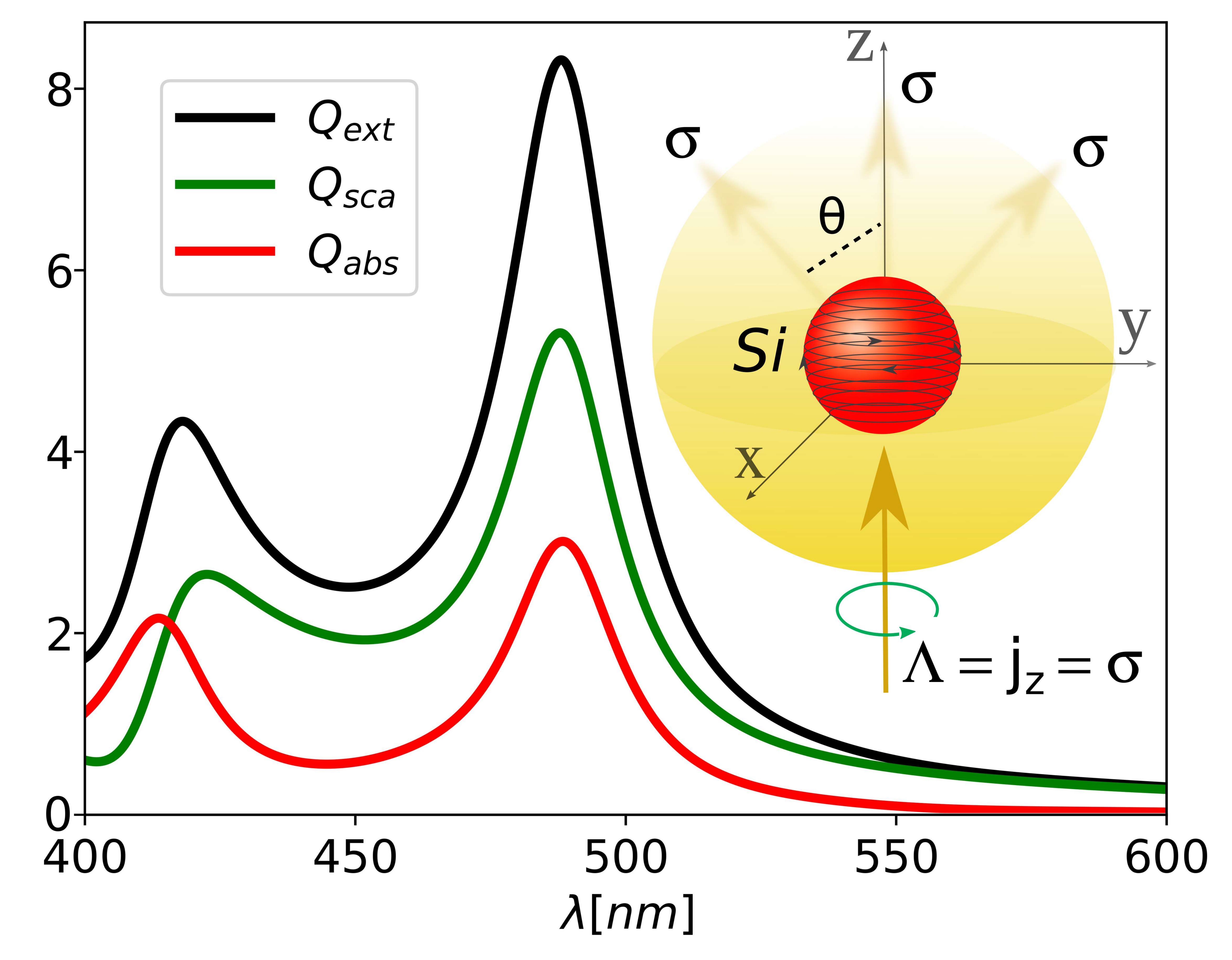}\captionsetup{justification= raggedright}
\caption{Theoretical extinction, scattering and absorption cross sections computed from Mie theory \citep{bohren2008absorption} for a 55 nm Si sphere excited by a plane wave with well-defined helicity, $j_z = \sigma$, in the visible spectral range. The dielectric function for Si were obtained from Ref.\cite{aspnes1983dielectric}. The absorbed light gives rise to a radiation torque, represented by the spinning arrows around the Si sphere. }\label{1} 
\end{figure}

The connection of the DoCP, expressible as shown in Stokes parameters, with SOI effects is given by the following relation
with the SAM per photon after scattering in the far-field. 
For a system with axial symmetry, as the one sketched in Fig. \ref{1}, the SAM and OAM per photon contributions  can be expressed as,
\begin{align}\label{SO:3}
s_z = -i\frac{\left({{\bf{E}}_\sigma^{\rm{scat*}}}  \times  {\bf{E}}^{\rm{scat}}_\sigma \right) \cdot \hat{\bf{z}}}{{{\bf{E}}_\sigma^{\rm{scat}}}^* \cdot {\bf{E}}^{\rm{scat}}_\sigma} = \Lambda \cos \theta,
\end{align}
\begin{equation}\label{SO:4}
\ell_z = \frac{{{\bf{E}}_\sigma^{\rm{scat*}}} \cdot \left( {{L}}_z  {\bf{E}}^{\rm{scat}}_\sigma \right)}{{{\bf{E}}_\sigma^{\rm{scat}}}^* \cdot {\bf{E}}^{\rm{scat}}_\sigma} =  \sigma - \Lambda \cos \theta ,
\end{equation}
where ${{L}}_z = -i\left( {\bf{r}} \times  {\bf{\nabla}} \right)_z =  -i  \partial_\varphi$ is the { OAM} operator in the incident direction. 

It is important to notice that the relations given so far are perfectly valid in the presence of absorption, since they do not depend on  energy conservation. For absorptive particles, the total (integrated) AM  is not a conserved quantity given that the particle will absorb a fraction of the incident photons. These optical losses give rise to a non-zero absorption cross section in the visible spectral range \cite{kuznetsov2012magnetic}, as it is illustrated in FIG. \ref{1}, where a plane wave with both well-defined helicity and AM per photon in the incident direction, $ j_z = \sigma =\pm 1$, impinges on an absorbing Si sphere of radius $a = 55$ nm. This non-zero absorption cross section is associated with the induction of a net optical torque to the target \citep{nieto2015optical}. Nonetheless, the scattered photons must preserve the incident AM per photon in the direction of incidence  as a simple consequence of the scatterer axial symmetry. This {is also illustrated in FIG. \ref{1}}. 


This conservation leads us to the study of the exchange between the SAM and OAM contributions per photon, which can be easily computed by inserting the DoCP after scattering \citep{olmos2018asymmetry}, i.e. Eq. \eqref{SO:1} into Eqs. \eqref{SO:3}-\eqref{SO:4}. In the electric and magnetic dipolar regime\citep{olmos2019enhanced}, 
\begin{align}\label{SO:5}
s_z = \frac{ 2  \sigma \cos \theta \left(\left(1 + \cos^2 \theta  \right)g  +  \cos \theta \right)}{{1 + \cos^2 \theta}  +4 g  \cos \theta },
\end{align} 
\begin{align}\label{SO:6}
\ell_z =\sigma \frac{\sin^2 \theta \left(1 + 2g \cos \theta \right)}{1 + \cos^2 \theta +4g \cos \theta},
\end{align} 
where
\begin{align}\label{SO:7}
g =  \frac{\text{Re}\left\{  \alpha_{\rm{E}}  \alpha^*_{\rm{M}}\right\}}{|\alpha_{\rm{E}}|^2 +  |\alpha_{\rm{M}}|^2} 
\end{align}
is the asymmetry parameter \citep{bohren2008absorption, gomez2012negative}.

It is worth mentioning that, recently, a bi-univocal relationship between the measurable DoCP and the $g$-parameter was found \citep{olmos2018asymmetry}, specifying that $\langle \Lambda \rangle = \Lambda_{\pi/2} = 2\sigma g$, where $\Lambda_{\pi/2}$ is the DoCP at the perpendicular direction to the incoming wave. This implies that the $g$-parameter is a measurable magnitude in the electric and magnetic dipolar regime, since it can be expressed in terms of the measurable $V$ and $I$ Stokes parameters (see Eq. \eqref{SO:2}).

It must be notice that even though Eqs. \eqref{SO:5}-\eqref{SO:6} imply that the SOI of light densities depend on the absorption via the asymmetry parameter, 
their expected values do not have this behaviour, since
\begin{align}\label{SO:8}
\langle S_z \rangle = \frac{ \int_\Omega 2  \sigma \cos \theta \left(\left(1 + \cos^2 \theta  \right)g  +  \cos \theta \right)  d \Omega}{ \int_\Omega \left(1 + \cos^2 \theta +4g \cos \theta \right) d \Omega} = \frac{\sigma}{2},
\end{align}
\begin{align}\label{SO:9}
\langle L_z \rangle  = \sigma \frac{\int_\Omega \sin^2 \theta \left(1 + 2g \cos \theta \right) d \Omega}{ \int_\Omega \left(1 + \cos^2 \theta +4g \cos \theta \right) d \Omega} = \frac{\sigma}{2},
\end{align}
with
\begin{equation}
\langle S_z \rangle  + \langle L_z \rangle  = \langle J_z \rangle = \sigma.
\end{equation}
These results agree with those obtained by Crichton and Marston in Ref. \citep{crichton2000measurable}. Although the absorption, along with any optical effects contained in the $g$-parameter, does not contribute to the expectation values of both OAM and SAM contributions in the dipolar regime, as we are going to analyze now, intriguing effects associated to the SOI of light may appear in absorbing targets

\begin{figure}[h]
\includegraphics[width = 1\columnwidth]{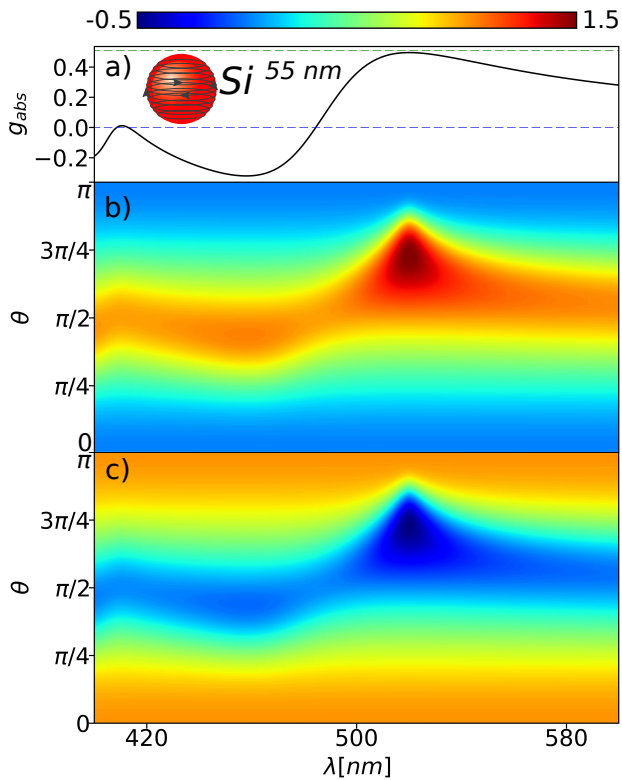}
\captionsetup{justification= raggedright}
\caption{(a) Asymmetry parameter, $g$, as a function of the incident wavelength, $\lambda$, in the visible spectral range, for a 55 nm Si sphere in the electric and magnetic dipolar regime. The maximum theoretical value (green dashed line), $g = 1/2$, given by the first Kerker condition, is denied by absorptive effects. (b)-(c) distributions of OAM and SAM per photon, respectively, as a function of the scattering angle, $\theta$, and wavelength $\lambda$. As indicated by the color-scale, warm colors (orange to red) illustrate values of OAM per photon that exceed the incident AM per photon, $\ell_z > j_z$. This effect is associated to negative spin density values, corresponding to blue regions in (c), since as the AM per photon in the incident direction is conserved, and $j_z = \ell_z + s_z = 1$ must be guaranteed.}\label{2} 
\end{figure}

\section{Discussion}

\begin{figure}[]
\includegraphics[width=1\columnwidth]{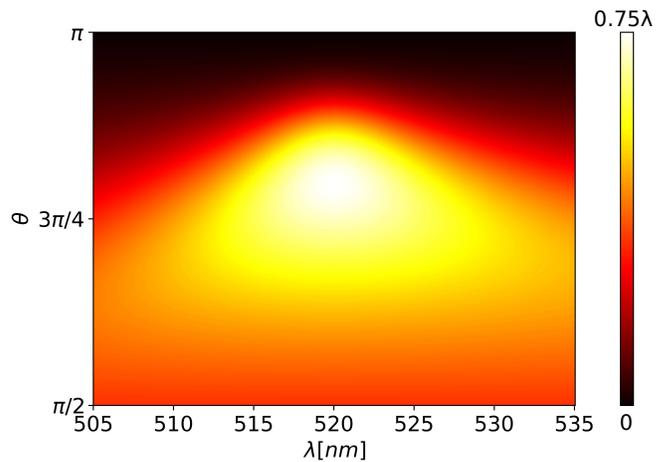}\captionsetup{justification= raggedright}
\caption{Spinning optical mirage, $\Delta/ \lambda$, calculated for a 55 nm Si sphere (see FIG.\ref{1}), as a function of $\lambda$ and $\theta$. The optical mirage reaches its maximum value at angles close to back-scattering ($\theta > 3\pi/4$), surpassing the maximum achivable values for pure electric (or magnetic) scatters.}\label{3} 
\end{figure}

Figure \ref{2} summarizes the re-distribution of the AM in OAM and SAM contributions per photon for the scattering problem described in FIG. \ref{1}. To this end, the incoming helicity is assumed  to be $\sigma =1$, which  corresponds with a left-circularly polarized plane wave. 
Figure \ref{2}(a) illustrates the $g$-parameter as a function of the incident wavelength, $\lambda$. As it can be seen, the maximum value, $g = 1/2$, is not reached due to the fact that the absorption denies the first Kerker condition,  $\alpha_{\rm{E}} \neq \alpha_{\rm{M}}$. Nevertheless, as it tends to be preserved,
$\Lambda \lesssim 1  \Longleftrightarrow g \lesssim 0.5$,
the OAM per photon (FIG.\ref{2}(b), {reaches} values that exceed the incident AM per photon in the incident direction, $\ell_z > j_z$. This phenomenon, previously referred as ``super-momentum" for absorptionless spheres \citep{bliokh2015spin, bliokh2017optical, araneda2018wavelength}, emerges as a simple consequence of the conservation of the AM per photon. According to Eq.\eqref{3}, $-1<s_z<1$, and since the AM per photon must be conserved after scattering, the OAM per photon can acquire values larger than the incident AM per photon, because $\ell_z + s_z = 1$.

It is worth mentioning that the ``super-momentum'' regime cannot be achieved for scatterers described by a single polarizability excited by a circularly polarized plane-wave, since in such scenarios $g=0$, and the OAM per photon is limited to $0<\ell_z<1$. This can be checked making $g=0$ in Eq.\eqref{SO:6}. Nevertheless, recently it has been demonstrated that elliptically polarized excitations, when focused by small numerical apertures can induce ``super-momentum'' on these kind of scatterers \citep{araneda2018wavelength}. 

Figure \ref{3} represents the color map of the apparent displacement associated to the scattering problem analyzed in FIG.\ref{1}, as a function of the scattering angle, $\theta$, and the incident wavelength, $\lambda$. In the ``super-momentum'' regime ($\ell_z >j_z$ {in Fig. \ref{2}}), the optical mirage doubles the values obtained from {a pure electric (or magnetic) scatterer} \citep{arnoldus2008subwavelength}, even when the absorption represent a third part of the extinction cross section. 
This can be easily understood from a simple relation \citep{olmos2019enhanced},
\begin{equation}
\frac{\Delta}{\lambda} = \frac{\ell_z}{\pi \sin \theta},
\end{equation} 
that relates the OAM per photon and the optical mirages in the dipolar regime.
Therefore, even though some of the incident light is absorbed by the spherical target, which in turn induces an optical torque, optical mirages persist. This shift can be comparable to the incident wavelength. Moreover, as a consequence of the interplay between the electric and magnetic dipoles, this optical mirage surpass the pure dipolar case shift \citep{arnoldus2008subwavelength, haefner2009spin}.

\section{Summary}

In conclusion, we have shown that absorption, which is included in the asymmetry parameter in the electric and magnetic dipolar regime, gives rise to an effective SAM to OAM exchange of the scattered photons. Based on the conservation of the incident AM per photon (as a direct consequence of the scatterer axial symmetry) we have shown that the so-called ``super-momentum'' regime, $\ell_z >j_z$,  emerges naturally in the scattering of an absorbing Si sphere. Consequently, a striking spinning optical mirage is found, its maximum value being comparable with the incident wavelength in the visible spectral range. Our results are discussed in terms of the DoCP and they could be experimentally refuted by far-field polarization measurements \cite{rodriguez2010optical,nechayev2018weak,neugebauer2018weak}.

\acknowledgments
This research was supported by the Spanish Ministerio de Econom\'ia y Competitividad (MICINN) and European Regional Development Fund (ERDF) Projects (FIS2015-69295-C3, FIS2016-80174-P, FIS2017-82804-P, FIS2017-91413-EXP) and FPU PhD Fellowship (FPU15/03566),
and by the Basque Government Projects (PI-2016-1-0041, KK- 2017/00089) and   PhD Fellowship (PRE-2018-2-0252). AGE received funding from the Fellows Gipuzkoa fellowship of the Gipuzkoako Foru Aldundia through FEDER ``Una Manera de hacer Europa''.

\bibliography{New_era}
\end{document}